\documentstyle{article}
\title{Quantum propagators in a random metric}
\author{Z. Brzezniak and Z. Haba\thanks{On leave 
of absence from Institute of Theoretical Physics,
University of Wroclaw, Plac Maxa Borna 9,  Wroclaw, Poland 
 e-mail:zhab@ift.uni.wroc.pl } \\ \\
 Department 
of Mathematics, University of Hull, \\  Hull HU6 7RX, UK \\
 e-mail:z.brzezniak@maths.hull.ac.uk}

\date{}
\begin{document}
\maketitle
\begin{abstract}
We consider  second order differential operators with 
coefficients which are Gaussian random fields.
 When the covariance becomes
singular at short distances then the propagators of the
Schr\"odinger equation as well as of the wave
equation behave in an anomalous way. In particular,
the Feynman propagator for the wave equation
is less singular than the one with deterministic coefficients.
We suggest some applications to quantum gravity.

PACS numbers: 03.65.-w,04.60.-m, 
\end{abstract}

Diffusions  generated  by  second
order differential operators with random coefficients have
been studied for some time.
Randomness changes the long time behaviour of fluctuations.
 However, the
short time behaviour is extremely stable with
respect to perturbations.
As an example, in a regular random field the mean square 
displacement for a small time is always linear in time.
 As a consequence, the mean value of the 
Green's function (the resolvent) in $d$-dimensional space
behaves as $\vert {\bf x}-{\bf y}\vert^{2-d}$ when the distance
$\vert {\bf x}-{\bf y}\vert$ tends to zero. Solutions
of the Schr\"odinger equation 
can usually be obtained from the solutions of the diffusion
equation by an analytic continuation in time. Hence,
an analogous behaviour is expected.

A long time ago it has been suggested
by Pauli ( see \cite{pauli}-\cite{deser})
 that quantization of the gravitational field can remove
divergencies of the conventional quantum field theory.
The divergencies arise as a consequence of the singular behaviour
of the Green's functions. Hence, if the Green's functions
averaged over the metric fluctuations were regular 
then the divergencies of the
quantum field theory would not appear at all.
Some perturbative calculations have been performed
recently, see \cite{ford} -\cite{ohanian}, showing that there
are no singularities on the light cone.

In this letter we study a simple model with a random
Gaussian metric. We discuss first the Schr\"odinger equation
with a random Hamiltonian. We consider a singular metric.
After a renormalization the mean value of the Feynman propagator 
of
the Schr\"odinger equation shows an anomalous behaviour
for a small time. Then, by means of the proper time method
\cite{feynman} we can obtain the Green's function for
the wave operator ( the Feynman's causal propagator).
We show that (if $d\leq 4)$ the expectation values of the Green's 
functions
are more regular at short distances than the ones with a
deterministic metric
(however our exact results do not confirm heuristic arguments
in refs.\cite{ford}-\cite{ohanian} which are based on approximate
calculations).

We consider the Schr\"odinger equation in $R^{2n}$
\begin{equation}
i\partial_{\tau}\psi=H\psi=-\frac{1}{2}\sum_{k=1}^{2n}
:\alpha_{k}^{2}:\partial_{k}^{2}\psi
\end{equation}
where $\alpha_{k}$ are independent homogeneous Gaussian 
random fields
(depending only on one coordinate of the vector ${\bf x}\in R^{2n}$)
with the same covariance
\begin{equation}
\langle \alpha_{k}(x) \alpha_{k}(y)\rangle=G(x-y)
\end{equation}
The Hamiltonian H in (1) is symmetric in $L^{2}(d{\bf x})$
because we choose $\alpha_{k}$ independent of $x_{k}$.
We begin with a regularized field $\alpha$ with a regular
covariance $G$. Then both terms in $:\alpha^{2}:(x)=
\alpha^{2}(x)-\langle \alpha^{2}(x)\rangle=\alpha^{2}(x)-
G(0)$ are well-defined. Subsequently, the Wick square
of a generalized random field $\alpha$ is defined as a limit
when the regularization is removed.

We solve the Schr\"odinger equation (1)
with the initial condition $\psi$  by means of a probabilistic 
version of the Feynman integral \cite{ha1}-\cite{ha2}
\begin{equation}
\psi_{\tau}({\bf x})=E[\psi\left({\bf q}_{\tau}\left({\bf x}
\right)\right)]
\end{equation}
where $E[..]$ is an expectation value with respect to
the Brownian motion defined as the Gaussian process on
$R^{2n}$ with the covariance
\begin{displaymath}
E[b_{j}(t)b_{k}(s)]=\delta_{jk}\min(s,t)
\end{displaymath}
In order to write the solution of the Schr\"odinger
equation (1) in the form (3) we assume that
$\psi$ is an analytic function. Then, we begin with
regularized random fields which are  analytic functions
as well.
 We remove the regularization only after a calculation
of the expectation values over $\alpha$.
The process ${\bf q}_{\tau}({\bf x})$ starts from ${\bf x}$
at $\tau=0$. It is  a solution of a set of stochastic equations
\cite{ike}.
In order to write down the solution explicitly we set
$\alpha_{2k}=1$ and $\alpha_{2k-1}({\bf x})=\alpha_{2k-1}(x_{2k})$
 for $k=1,..,n$. Then,
\begin{equation}
q_{2k}(\tau,{\bf x})=x_{2k}+\lambda b_{2k}(\tau)
\end{equation}
and the odd components of the process can be expressed by
the Ito integral
\begin{equation}
q_{2k-1}(\tau,{\bf x})=x_{2k-1}+\lambda\int_{0}^{\tau}
\alpha_{2k-1}\left(q_{2k}\left(s,{\bf x}\right)\right)db_{2k-1}(s)
\end{equation}
where  $\tau\geq 0$ and
\begin{displaymath}
\lambda=\sqrt{i}=\frac{1}{\sqrt{2}}(1+i)
\end{displaymath}
It can be checked by direct differentiation using the Ito
calculus \cite{ike} that $\psi_{\tau}$ solves the
Schr\"odinger equation (1) for each (regularized) $\alpha$.
If we represent $\psi$ in terms of its Fourier transform then
we can calculate explicitly the average over $\alpha$ 
of the products
\begin{displaymath}
\langle\psi_{\tau_{1}}({\bf x}_{1})...\overline{\psi}_{s_{1}}
({\bf y}_{1})...\rangle
\end{displaymath}
We restrict ourselves to the expectation value of the Feynman 
propagator
$K_{\tau}({\bf x},{\bf y})$. The time evolution
is determined by the propagator
\begin{displaymath}
\psi_{\tau}({\bf x})=\int K_{\tau}({\bf x},{\bf y})\psi({\bf y})
d{\bf y}
\end{displaymath}
 We further specify the random fields
$\alpha_{k}$ in eq.(1) by assuming that their covariance is of the 
form
\begin{equation}
G(x)=\int_{-\infty}^{\infty}d\nu\rho(\nu)\cos(\nu x^{4})
\end{equation}
Then, $G(\sqrt{i}x)=G(x)$.
For the computation of $\langle K_{\tau}\rangle$ it is sufficient
 to insert into eq.(3) $\exp(-i{\bf py})$
as the Fourier transform of $\psi({\bf x})=\delta({\bf x}-{\bf y}) $ and
to note that the Fourier transform of the r.h.s. of eq.(3)
is $\exp(-i{\bf py}-i{\bf px}+i{\bf p}{\bf q}_{\tau}({\bf x}))$.
The computation of the Gaussian integral over $\alpha$
and the momentum integrals $p_{2k-1}$ is elementary.
As a result we obtain the following formula
\begin{equation}
\begin{array}{l}
\langle K_{\tau}({\bf x},{\bf y})\rangle=(2\pi)^{-n}E[\int 
\prod_{k=1}^{n} dp_{2k}
\exp\left(ip_{2k}\left(x_{2k}-y_{2k}\right)
+i\lambda p_{2k}b_{2k}\left(\tau\right)\right) 
\cr
\left(2\pi\Gamma_{k}\left(\tau\right)\right)^{-\frac{1}{2}}
\exp\left(\frac{i}{2}
\left(x_{2k-1}-y_{2k-1}\right)^{2}/\Gamma_{k}(\tau)\right)]
\end{array}
\end{equation}
where 
\begin{equation}
\Gamma_{k}(\tau)=2\int_{0}^{\tau}db_{2k-1}(t)\int_{0}^{t}db_{2k-1}(s)
G\left(b_{2k}\left(t\right)-b_{2k}\left(s\right)\right)
\end{equation}
In eq.(7) we have considered first a regularized $G$. The Gaussian
integral over $\alpha $ gives the double integral $\int dbdb G$.
Then, we obtain $\Gamma=\int dbdbG-G(0)\tau$ in eq.(7)
 ($G(0)$ comes from the Wick square in eq.(1)).
The replacement of the double (unordered) integral by a time-
ordered
one leads to the formula (8).

We can calculate explicitly some "moments" of $K$, e.g.,
\begin{equation}
\langle x_{2k-1}^{2r}\rangle=\int d{\bf y} y_{2k-1}^{2r}
\langle K_{\tau}(0,{\bf y})\rangle
=E[ \Gamma_{k}^{2r}]
\end{equation}
Under quite general assumptions the kernel $K_{\tau}$ of a regular 
second order
differential operator behaves for a small time as
 $\tau^{-n}\exp(\frac{i}{\tau}A({\bf x})\vert {\bf x}-{\bf y}\vert^{2})$.
 Hence, for a small time
$\langle x^{2r}\rangle\simeq \tau^{r}$. We show that such a 
behaviour 
fails in a singular random field.

In order to simplify the subsequent discussion we choose 
$\rho(\nu)$ in 
$G$ (eq.(6)) in a
scale invariant form
\begin{displaymath}
\rho(\nu)=const \vert \nu\vert^{\frac{\gamma}{2}-1}
\end{displaymath}
where $\gamma<\frac{1}{2}$ if the stochastic integrals in eq.(8) are 
to make
sense. Then
\begin{equation}
G(x)=\vert x\vert^{-2\gamma}
\end{equation}
The integral (9) could be calculated explicitly but this is not
necessary. We can use the scale invariance of the Brownian motion
(saying that $\sqrt{c} b(t/c)$ has the same probability
 distribution as $b(t)$)
in order to conclude that for an integer $r$
\begin{equation}
\langle x_{2k-1}^{2r}\rangle =A_{r}\tau^{r(1-\gamma)}
\end{equation}
with  certain constants $A_{r}$ (here $A_{1}=0$).
The even coordinates are described by a simple formula for
the "moments" 
$\langle x_{2k}^{2r}\rangle =E[\left(b_{2k}\left(\tau\right)\right)^{2r}]
=C_{r}\tau^{r}$.

We did not use an assumption that $\alpha_{k}^{2}$ in eq.(1) have 
the same
sign , i.e., we could take  some of $\alpha_{k}=i\tilde{\alpha}_{k}$ 
where
$\tilde{\alpha}_{k}$ has the covariance (2). Let us assume that
 $\alpha_{1}$
is purely imaginary, then $\Gamma_{1}\rightarrow - \Gamma_{1}$ 
in eq.(7).
We obtain on the r.h.s. of eq.(1) the operator 
( which is symmetric in $L^{2}(dx)$)
\begin{equation}
{\cal A}=\frac{1}{2}g^{\mu\nu}\partial_{\mu}
\partial_{\nu}
\end{equation}
 with the  metric $g$ which has  the Minkowski signature
$(\eta_{\mu\nu})=(1,-1,-1,..,-1) $. This is the ghost 
operator in quantum gravity
(in a particular gauge). We can obtain the causal Feynman 
propagator
of ${\cal A}$  (the Green's function) by means of the proper 
time method \cite{feynman}
\begin{equation}
{\cal A}^{-1}(x,y)=i\int_{0}^{\infty} d\tau\left(\exp\left(-i\tau{\cal
A}\right)\right)\left(x,y\right)
\end{equation} 
The Green's function depends only on $x-y$. It cannot be 
calculated
explicitly. However, we obtain a simple formula if we integrate over
the even coordinates
\begin{equation}
\begin{array}{l}
\int \prod_{k}dx_{2k}\langle{\cal A}^{-1}(x,y)\rangle 
=i\int_{0}^{\infty}d\tau
\cr
E[\prod_{k}\left(2\pi\Gamma_{k}\left(\tau)\right)\right)^{-\frac{1}{2}}
\exp\left((-1)^{e_{k}}\frac{i}{2}\left(x_{2k-1}-y_{2k-
1}\right)^{2}/\Gamma_{k}
\left(\tau\right)\right)]
 \end{array}
\end{equation}
where $(-1)^{e_{k}}=\pm 1$  is the signature of the metric in eq.(12).
Let us denote
\begin{displaymath}
\left(x-y\right)^{2}=\sum_{k=1}^{n}
\left((-1)^{e_{k}}\left(x_{2k-1}-y_{2k-1}\right)^{2}
      +\left(x_{2k}-y_{2k}\right)^{2}\right)
      \equiv (x-y)_{o}^{2}+(x-y)_{e}^{2}
\end{displaymath}
Then, changing the integration variable $\tau \rightarrow
((x-y)_{o}^{2})^{\frac{1}{1-\gamma}}\tau$ we can conclude that
at short distances
\begin{equation}
\int \prod_{k}dx_{2k}\langle {\cal A}^{-1}(x,y)\rangle\simeq
((x-y)_{o}^{2})^{-\frac{n}{2}+\frac{1}{1-\gamma}}
\end{equation}
The integral (15) is less singular than the one in the
deterministic case which corresponds to $\gamma=0$.

An estimate of the Green's function  $<{\cal A}^{-1}(x,y)>$
when $x\rightarrow y$ 
is simple if $x\rightarrow y$ either on the hyperplane
$(x-y)_{e}^{2}=0$ or on the one with $(x-y)_{o}^{2}=0$.
In eq.(7) we can write $b_{2k}(\tau)=\sqrt{\tau}b_{2k}(1)$
and change the integration variable 
$\sqrt{\tau}p_{2k}=p_{2k}^{\prime}$.
Then, in the first case by rescaling
$\tau\rightarrow \left(\left(x-y\right)_{o}^{2}
\right)^{\frac{1}{1-\gamma}}\tau$
we obtain for short distances
\begin{equation}
\langle {\cal A}^{-1}(x,y)\rangle
\simeq \left(\left(x-y\right)_{o}^{2}\right)
^{-\frac{n}{2}+(1-\frac{n}{2})/(1-\gamma)}
\end{equation}
In the case $(x-y)_{o}^{2}=0$ the rescaling $\tau \rightarrow
(x-y)_{e}^{2}\tau$ gives the result
\begin{equation}
\langle {\cal A}^{-1}(x,y)\rangle
\simeq \left(\left(x-y\right)_{e}^{2}\right)^{-\frac{n}{2}
+1-\frac{n}{2}(1-\gamma)}
\end{equation}
Let us still consider the example
when $d=2n=3$ and $\alpha_{2}=\alpha_{3}=1$ in eq.(1).
Then, the propagator behaves
as $\vert x_{1}-y_{1}\vert^{-1}$ for $x_{2}-y_{2}=
x_{3}-y_{3}=0$ and if $x_{1}=y_{1}$ then the short distance
behaviour is $\vert {\bf x}-{\bf y}\vert^{\gamma-1}$.
We conclude that (if $d\leq 4$)  increasing singularity
of $G$ leads to increasing regularity of the Feynman's causal
propagator. However, if $\gamma<\frac{1}{2}$ then except of
 the case $d=2n=2$, the propagators remain singular
 ( contrary to the suggestion of refs.
\cite{ford}-\cite{ohanian}). By a formal argument,
when $d=2n=4$, we can set $\gamma=1$ in
eqs.(7) and (17) with a conclusion that the singularity
disappears. However,
the case $\gamma=1$ requires a careful examination because the 
integral
(8) becomes divergent.
For the general metric $ g_{\mu\nu}$ in eq.(12)
we can perform only perturbative calculations. We
put $g_{\mu\nu}=\eta_{\mu\nu}+h_{\mu\nu}$. We apply the
stochastic representation (3) solving the corresponding
stochastic equations for $q_{\tau}$ (see \cite{ike}\cite{ha2})
till the first order in $h$. We assume that $h$ is Gaussian
with the singularity $(x^{2})^{-\gamma}$ on the light cone.
These assumptions give the short
distance behaviour $((x-y)^{2})^{-n+1+\frac{1}{2}(2-n)\gamma/(1-
\gamma)}$.
It agrees with eq.(16), i.e.,
when all the coordinates relevant for the asymptotics
are disturbed by a fluctuating metric.

In quantum field theory we obtain an average of 
many Green's functions. This average value has
a  functional representation similar to (7) (there will be
a larger number of proper times $\tau$ ). We can obtain the same 
conclusion
that the average is less singular than in quantum field theory
with a fixed metric.
\vspace{2cm}

This research is partially supported by The Royal Society

\end{document}